\documentstyle[preprint,aps,version2]{revtex}
\begin{document}
\draft
\preprint{\vbox{\baselineskip=12pt{\hbox{CALT-68-1867}\hbox{hep-th/9306006}}}}
\begin{title}
Non-abelian vortices and non-abelian statistics
\end{title}
\author{Hoi-Kwong Lo and John Preskill}
\begin{instit}
California Institute of Technology, Pasadena, CA 91125
\end{instit}
\begin{abstract}
We study the interactions of non-abelian vortices in two spatial dimensions.
These interactions have novel features, because the Aharonov-Bohm effect
enables a pair of vortices to exchange quantum numbers.  The cross section for
vortex-vortex scattering is typically a multi-valued function of the scattering
angle.  There can be an exchange contribution to the vortex-vortex scattering
amplitude that adds coherently with the direct amplitude, even if the two
vortices have distinct quantum numbers.  Thus two vortices can be
``indistinguishable'' even though they are not the same.
\end{abstract}

\narrowtext
\section{INTRODUCTION}
\label{sec:intro}
It is well known that exotic generalizations of fermion and boson statistics
are possible in two spatial dimensions.   The simplest, and most familiar, such
generalization is anyon statistics \cite{leinaas,wilczek82}.  When two
indistinguishable anyons are
adiabatically interchanged (or one anyon is rotated by $2\pi$), the many-body
wave function acquires the phase $e^{i\theta}$, where $\theta$ can take any
value.  An instructive example of an object that obeys anyon statistics is a
composite of a magnetic vortex (with magnetic flux $\Phi$) and a charged
particle (with charge $q$) \cite{wilczek82}.  Then the anyon phase arises as a
consequence of
the Aharonov-Bohm effect, with $e^{i\theta}=e^{iq\Phi}$.  Furthermore, anyon
statistics is actually known to be realized in nature, in systems that exhibit
the fractional quantum hall effect \cite{fqhe}.

It is natural to consider a further generalization:  non-abelian
statistics \cite{goldin85,bala,wilczekwu,bucher}.  A
particular type of non-abelian statistics is realized by the non-abelian
vortices (and vortex-charge composites) that occur in some spontaneously broken
gauge theories.  Loosely speaking, the unusual feature of the many-body physics
in this case is that the quantum numbers of an object depend on its {\it
history}.  In particular, if one vortex is adiabatically carried around
another, the quantum numbers of both may change, due to a non-abelian variant
of the Aharonov-Bohm effect.  Thus, whether two bodies are identical is not a
globally defined notion.

There is no firm observational evidence for the existence of objects that obey
this type of quantum statistics.  Perhaps such objects will eventually be found
in suitable condensed matter systems.  (Analogous non-abelian defects
associated with spontaneous breakdown of {\it global} symmetries are observed
in liquid crystals \cite{mermin} and ${}^3 He$ \cite{volovik}.)
In any event, the physics of non-abelian
vortices is intrinsically interesting and instructive.  For one thing, it
forces us to carefully consider some subtle aspects of non-abelian gauge
invariance.

In this paper, we will focus on the Aharonov-Bohm interactions of a pair of
non-abelian vortices.  This is, of course, much simpler and much less
interesting than the problem of three or more bodies.  Nevertheless, an
important conceptual point will be illuminated by our calculation of the
vortex-vortex scattering cross section.  We will see that this cross section is
in general multi-valued.  While we have learned to be undisturbed, at least in
certain contexts, by multi-valued wave functions, a cross section is directly
observable, and so is ordinarily expected to be a single-valued function of the
scattering angle.  But the multi-valuedness of the cross section for
vortex-vortex scattering follows naturally from the ambiguity in assigning
quantum numbers to the vortices.

Indeed, multi-valued scattering cross sections are a generic consequence of the
non-abelian Aharonov-Bohm effect---they arise in the scattering of a charge off
a vortex as well.  It is useful to consider the case of the ``Alice''
vortex \cite{schwarz,alfordbenson,preskillkrauss,bucherlopreskill},
which has the property that when a positively charged particle is adiabatically
transported around the vortex, it becomes negatively charged.  When a
positively charged particle scatters from an Alice vortex, the scattered
particle may be either positively charged or negatively charged.  Thus there
are two measurable exclusive cross sections,\footnote{Strictly speaking, these
cross sections do not exist, because there are no asymptotic charged particle
states in two-dimensional electrodynamics; see Sec.~\ref{sec:alice} for further
discussion.} $\sigma_+(\theta)$ and
$\sigma_-(\theta)$.  Though the inclusive cross section $\sigma_{\rm
inc}=\sigma_+
+ \sigma_-$ is single valued, the exclusive cross sections are not; they are
double-valued and obey the conditions
\begin{equation}
\sigma_+(\theta + 2\pi)=\sigma_-(\theta)\; ,\quad\quad
\sigma_-(\theta + 2\pi)=\sigma_+(\theta)\; .
\label{eq:alicecrossmono}
\end{equation}
The double-valuedness of the exclusive cross sections is an unavoidable
consequence of the feature that a charged particle that voyages around the
Alice vortex returns to its starting point with its charge flipped in sign.  We
might imagine measuring the $\theta$-dependence of the cross section by
gradually transporting a particle detector around the scattering center.  But
then a detector that has been designed to respond to positively charged
particles will have become a detector that responds to negatively charged
particles when it returns to its starting point.  Alternatively, we might catch
the scattered particle, and then carry it back along a specified path to a
central laboratory for analysis.  But then the outcome of the analysis will
depend upon the path taken.  While we may (arbitrarily) associate a definite
path with each value of the scattering angle, this path cannot vary
continuously with $\theta$. A convention for choosing the path artificially
restricts the exclusive cross
sections to a single branch of the two-valued function, and introduces a
discontinuity in the measured cross sections.  As we will discuss in more
detail below, the cross sections for non-abelian vortex-vortex scattering have
similar multi-valuedness properties.

In the case of vortex-vortex scattering (unlike the case of scattering a
charged particle off of a vortex), effects of quantum statistics can be
exhibited.  That is, there may be an exchange contribution to the scattering
amplitude that interferes with the direct amplitude.  The existence of an
exchange contribution means that the two vortices must be regarded as
indistinguishable particles---it is not possible in principle to keep track of
which vortex is which. The unusual feature of
non-abelian vortex-vortex scattering is that exchange scattering can occur even
if the initial vortices are objects with distinct quantum numbers.  The
vortices are different, yet they are indistinguishable.

Much that we will say in this paper has been anticipated elsewhere.  That the
quantum numbers of non-abelian vortices can not be globally defined was first
emphasized by Bais \cite{bais80}.  (The corresponding observation for
defects associated with a spontaneously broken {\it global} symmetry was
made earlier, by Po\'enaru and Toulouse \cite{toulouse}.)
Wilczek and Wu \cite{wilczekwu} and
Bucher \cite{bucher} discussed the implications for
vortex-vortex scattering.  E. Verlinde \cite{verlinde}
worked out a general formula for the
{\it inclusive} cross section in Aharonov-Bohm scattering, in terms of the
matrix
elements of the ``monodromy'' operator, and Bais {\it et al.}
\cite{bais92} developed a
powerful algebraic machinery that can be used to compute these matrix elements
(among other things).  The main new contributions here are a computation of the
{\it exclusive} cross sections for the various possible quantum numbers of the
final
state vortices, and an analysis of vortex-vortex scattering that incorporates
the exchange of ``indistinguishable'' vortices.  (Wilczek and Wu
\cite{wilczekwu} attempted to
calculate the exclusive cross
sections, but because they missed the multi-valuedness properties of these
cross sections, they did not obtain the correct answer.)  Once properly
formulated, the calculation of these exclusive cross sections is very closely
related to the analysis of scattering in (2+1)-dimensional gravity, which was
first worked out by 't Hooft \cite{hooft} and Deser and Jackiw
\cite{deser}.

The remainder of this paper is organized as follows:  In Sec.~II, we review how
the the quantum numbers of non-abelian vortices are modified by an exchange,
and we extend the discussion in Sec.~III to the case of vortices that also
carry charge.  We recall the general theory of the quantum mechanics of
indistinguishable particles in Sec.~IV, and describe how the special case of
non-abelian vortices fits into this general theory.  In Sec.~V, we calculate
the exclusive cross sections for non-abelian Aharonov-Bohm scattering of a
projectile off of a fixed target.  The case of vortex-vortex scattering is
analyzed in detail, and we emphasize and explain the multi-valuedness
properties of these cross sections.  The case of two-body scattering in the
center-of-mass frame is discussed in Sec.~VI.  This calculation includes the
contribution due to the exchange of ``indistinguishable'' vortices.  In
Sec.~VII, we extend the previous discussion to the case where the unbroken
gauge group is continuous, such as the case of the ``Alice'' vortex.  Sec.~VIII
contains our conclusions.

\section{NON-ABELIAN FLUX AND THE BRAID OPERATOR}
\label{sec:flux}

We consider, in two spatial dimensions, a gauge theory with underlying gauge
group $G$, which we may take to be connected and simply connected.  Suppose
that the gauge symmetry is spontaneously broken, and that the surviving
manifest gauge symmetry is $H$.  We will assume for now that $H$ is discrete
and
non-abelian.  The case of continuous $H$ will be briefly discussed in Sec.
\ref{sec:alice}.

This pattern of symmetry breaking will admit stable classical vortex solutions.
A vortex carries a ``flux'' that can be labeled by an element of the unbroken
group $H$.  To assign a group element to a vortex, we arbitrarily choose a
``basepoint'' $x_0$ and a path $C$, beginning and ending at $x_0$ that winds
around the vortex.  The effect of parallel transport in the gauge potential of
the vortex is then encoded in
\begin{equation}
a(C,x_0)=P\exp\left(i\int_{C,x_0} A\right)\in H(x_0)\; .
\end{equation}
This group element takes a value in the subgroup $H(x_0)$ of G that preserves
the Higgs condensate at the point $x_0$, since transport of the condensate
around the vortex must return it to its original value.  If $H$ is discrete,
then $a(C,x_0)$ will remain unchanged as the path $C$ is smoothly deformed, as
long as the path never crosses the cores of any vortices.  (The gauge
connection is locally flat outside the vortex cores, with curvature
singularities at the cores.)

The flux of a vortex can be measured via the Aharonov-Bohm effect
\cite{alfordmrwilczek,alfordcolemanmr}.  We can
imagine performing a double slit interference experiment with a beam of
particles that transform as some representation $R$ of $H$.  If we then repeat
the experiment with the vortex placed between the two slits, the change in
the interference pattern reveals
\begin{equation}
\langle u^{(R)}|D^{(R)}(a)|u^{(R)}\rangle\; ,
\end{equation}
where $|u^{(R)}\rangle$ is the internal wave function of the particles in the
beam.  (The shift in the interference fringes is determined by the phase of
this quantity, and the amplitude of the intensity modulation is determined by
its modulus.)  By measuring this for various $|u^{(R)}\rangle$'s, all matrix
elements
of $D^{(R)}(a)$ can be determined, and hence, if the representation is
faithful, $a$ itself.

However, the flux of the vortex is not a gauge-invariant quantity.  A gauge
transformation $h\in H(x_0)$ that preserves the Higgs condensate at the
basepoint transforms the flux according to
\begin{equation}
h:\quad a\to hah^{-1} \; .
\end{equation}
(This gauge transformation is just a relabeling of the particles that are used
to perform the measurement of the flux.)
Since the gauge transformations act transitively on the conjugacy class in $H$
to which the flux belongs, one might be tempted to say that the flux of a
vortex should really be labeled by a conjugacy class rather than a group
element.  But that is not correct.   If there are two vortices, labeled by
group
elements $a$ and $b$ with respect to the same basepoint $x_0$, then the effect
of a gauge transformation at $x_0$ is
\begin{equation}
h:\quad a\to hah^{-1} \; , \quad b\to hbh^{-1}\; .
\end{equation}
Thus, if $a$ and $b$ are distinct representatives of the same class, they
remain distinct in any gauge.

More generally, we can imagine assembling a ``vortex bureau of standards,''
where standard vortices corresponding to each group element are stored.  If a
vortex of unknown flux is found, we can carry it back to the bureau of
standards and determine which of the standard vortices it matches.
(Alternatively, we can find out which antivortex it annihilates.)  Thus, though
there is arbitrariness in how we assign group elements to the standard
vortices, once our standards are chosen there is no ambiguity in assigning a
label to the new vortex.

We might have said much the same thing about measuring the color of a quark.
Although the color is not a gauge-invariant quantity, we can erect a quark
bureau of standards in which standard red, yellow, and blue quarks are kept.
When a new quark is found, we can carry it back to the bureau and determine its
color relative to our standard basis.  However, in the case where there are
light gauge fields, curvature of the gauge connection is easily excited.  We
may find, then, that the outcome of the measurement  of the color depends on
the path that is chosen when the quark is transported back to the bureau.

In the case where the unbroken gauge group is discrete, there are no light
gauge fields.  The measurement of the flux of a vortex is unaffected by a
deformation of the path that is used to bring the vortex to the bureau of
standards, as long as the path does not cross the cores of any other vortices.
But when other vortices are present, there is a discrete choice of
topologically distinct paths, and the measured flux will in general depend on
how we choose to weave the vortex among the other vortices on the way back to
the bureau.  This ambiguity in measuring the flux is the origin of the
``holonomy interaction'' among vortices \cite{bais80}, and of Aharonov-Bohm
vortex-vortex
scattering \cite{wilczekwu,bucher}.

To characterize this interaction, we consider how the fluxes assigned to a pair
of vortices are modified when the two vortices are adiabatically interchanged,
as depicted in Fig.~\ref{figA}.  Here $\alpha$ and $\beta$ are standard paths,
beginning
and ending at the basepoint $x_0$, that are used to define the flux of the two
vortices; the corresponding group elements are $a$ and $b$ respectively.  When
the two vortices are interchanged (in a counterclockwise sense), these paths
can be dragged to new paths $\alpha'$ and $\beta'$, in such a way that no path
ever crosses any vortex.  Thus, the group elements associated with transport
along $\alpha'$ and $\beta'$ are, after the interchange, still $a$ and $b$
respectively.  But the final paths are topologically distinct from the initial
paths; from Fig.~\ref{figA}b we see that
\begin{equation}
\alpha'=\beta\rightarrow a\;, \quad \beta'=\beta^{-1}\alpha\beta\rightarrow b\;
{}.
\end{equation}
(Here, in order to be consistent with the rules for composing path-ordered
exponentials, we have chosen an ordering  convention in which $\alpha\beta$
denotes the path obtained by first traversing $\beta$, then $\alpha$.)
We conclude that, after the interchange, the effect of parallel transport
around $\alpha$ is given by the group element $aba^{-1}$.  The effect of the
interchange on the two vortex state can be expressed as the action of the {\it
braid operator} ${\cal R}$, where
\begin{equation}
\label{braidop}
{\cal R}:\quad |a,b\rangle\to |aba^{-1},a\rangle\; .
\end{equation}
Naturally, the braid operator preserves the ``total flux'' $ab$ that is
associated with counterclockwise transport around the vortex pair, for this
flux could be measured by a particle that is very far away from the pair, and
cannot be affected by the interchange.
If the interchange is performed twice (which is equivalent to transporting one
vortex in a counterclockwise sense about the other), the state transforms
according to
\begin{equation}
\label{monodromy}
{\cal R}^2:\quad |a,b\rangle\to |(ab)a(ab)^{-1},(ab)b(ab)^{-1}\rangle\; ;
\end{equation}
both fluxes are conjugated by their combined ``total flux'' $ab$.

This result has a clear, gauge-invariant meaning.  Suppose that two vortices
are carried from their initial positions to the vortex bureau of standards
along the paths shown in Fig.~\ref{figB}a, and are found to have fluxes $a$ and
$b$.  Then
if they are carried to the bureau along the alternative paths shown in
Fig.~\ref{figB}bc,
the outcome of the flux measurement will be different, as expressed in Eq.\
(\ref{monodromy}).

\section{FLUX-CHARGE COMPOSITES}
\label{sec:flux-charge}
The above discussion can be generalized to the case of objects that carry both
flux and charge.  But there is one noteworthy subtlety. The ``charge'' of an
object is defined by its transformation properties under global gauge
transformations.  If the object carries flux, however, there is a topological
obstruction to implementing the global gauge transformations that do not
commute with the flux \cite{nonreal,alfordbenson,preskillkrauss}.  If the flux
is $a$, only a subgroup of $H$, the
centralizer $N(a)$ of the flux, is ``globally realizable'' acting on the
vortex.  Thus, the charged states of a vortex with flux $a$ transform as a
representation of $N(a)$ rather than the full group $H$.

We can understand the physical meaning of this obstruction if we think about
measuring charge via the Aharonov-Bohm effect.  The charge can be measured in a
double slit interference experiment, by observing the effect on the
interference pattern when various vortices are placed between the slits.  But
if the particles in the beam carry flux $a$, and the vortex between the slits
carries flux $b$, then no interference pattern is seen if $a$ and $b$ do not
commute.  The trouble is that, due to the holonomy interaction, the objects
that pass through the respective slits carry different values of the flux when
they arrive at the detector, and so do not interfere. (See Fig.~\ref{figC}.)
Even more to the point,
the slit that the object passed through becomes correlated with the state of
the vortex that is placed between the slits, because both fluxes become
conjugated as in Eq.\ (\ref{monodromy}).  Thus, the superposition of particles
that passed through the two slits becomes incoherent, and there is no
interference.  There will be an interference pattern, and a successful charge
measurement, only if the flux between the slits commutes with the flux $a$
carried by the particles in the beam.  Hence only the transformation properties
under $N(a)$ can be measured.

Since the global gauge transformations that can be implemented actually commute
with the flux, a non-abelian vortex that carries charge behaves much like an
abelian flux-charge composite.  If the vortex carries flux $a$ and transforms
as an irreducible representation $(R^{(a)})$ of $N(a)$, then, since $a$ lies in
the
center of its centralizer $N(a)$, it is represented by a multiple of the
identity in $R^{(a)}$ (by Schur's lemma),
\begin{equation}
D^{(R^{(a)})}(a)=e^{i\theta_{R^{(a)}}}\, {\bf 1}_{R^{(a)}}\; .
\label{anyphase}
\end{equation}
Thus, the charged vortices are anyons, and $e^{i\theta_{R^{(a)}}}$ is the anyon
phase.
A spin-statistics connection holds for these anyons
\cite{wilczek82,frolich}, in the sense that an
adiabatic interchange of a pair is equivalent to rotating one by $2\pi$---we
have $e^{2\pi iJ}=e^{i\theta_{R^{(a)}}}$.

The non-abelian character of the vortices becomes manifest when we consider
combining together two flux-charge composites, and decomposing into states of
definite charge.  The decomposition has the form
\begin{equation}
|a,R^{(a)}\rangle\otimes |b,R^{(b)}\rangle=\oplus_R|ab, R^{(ab)}\rangle\; ,
\end{equation}
where $R^{(a)}$ denotes an irreducible representation of $N(a)$.  The
nontrivial problem of decomposing a direct product of a representation of
$N(a)$ and a representation of $N(b)$ into a direct sum of representations of
$N(ab)$ is elegantly solved by the representation theory of quasi-triangular
Hopf algebras, as described in the beautiful paper of Bais {\it et al.}
\cite{bais92} (see also \cite{orbifold,bantay}).  This
decomposition also diagonalizes the monodromy matrix ${\cal M}\equiv {\cal
R}^2$ that acts on the two vortex state when one vortex winds
(counterclockwise) around the other \cite{rehren,orbifold}:
\begin{equation}
{\cal M}\equiv {\cal R}^2=\exp\left[{i\left(\theta_{R^{(ab)}}-\theta_{R^{(a)}}
-\theta_{R^{(b)}}\right)}\right]\; .
\label{monodiag}\end{equation}
Eq.\ (\ref{monodiag}) follows from Eq.\ (\ref{anyphase}) and the
spin-statistics connection for anyons, for the action of the monodromy operator
is equivalent to a rotation of the vortex pair by $2\pi$, accompanied by a
rotation of each member of the pair by $2\pi$ in the opposite sense.

A remarkable property of this decomposition is that a pair of uncharged
vortices can be combined together to form an object that carries charge
\cite{alfordbenson,preskillkrauss,bais92}.  This
is called ``Cheshire charge,'' in homage to the Cheshire cat; the charge can be
detected via the Aharonov-Bohm interaction of the pair with another, distant,
vortex, but it cannot be localized anywhere on the vortex cores or in their
vicinity.
Charge can be transferred to or from a pair of vortices due to the
Aharonov-Bohm
interactions of the pair with another charged object that passes through the
two vortices \cite{preskilletc,bucherleepreskill,alfordcolemanmr}.  Since the
pair generically carries a fractional spin given by
$e^{2\pi i J}=e^{i\theta_{R^{(ab)}}}$, angular momentum is also transferred in
these processes \cite{imbo}.

\section{NON-ABELIAN QUANTUM STATISTICS}
In this Section, we will briefly describe how the non-abelian statistics obeyed
by non-abelian vortices fits into general discussions of quantum statistics
that have appeared in the literature.

In general discussions of the quantum statistics of indistinguishable
particles, the following framework is usually adopted:  Suppose that the
position of each particle takes values in a manifold $M$ (like $R^d$).  For $n$
{\it distinguishable} particles, we would take the classical configuration
space to be $M^n=M\times M \times \cdots \times M$.  For {\it
indistinguishable} particles (other than bosons), we must restrict the
positions so that no two particles coincide, and we must identify
configurations that differ by a permutation of the particles.  Thus, the
classical configuration space  becomes
\begin{equation}
{\cal C}_n=\left[M^n-D_n\right]/S_n\; ,
\end{equation}
where $D_n$ is the subset of $M^n$ in which two or more points coincide, and
$S_n$ is the group of permutations of $n$ objects.  In general, this
configuration space is not simply connected, $\pi_1({\cal C}_n)\ne 0$.

We may now imagine quantizing the theory by using, say, the path integral
method.  The histories that contribute to the amplitude for a specified initial
configuration to propagate to a specified final configuration divide up into
disjoint sectors labeled by the elements of $\pi_1({\cal C}_n)$.  We have the
freedom to weight the contributions from the different sectors with different
factors, as long as the amplitudes respect the principle of conservation of
probability.  In general, there are distinct choices for these weight factors,
which correspond to physically inequivalent ways of quantizing the classical
theory \cite{laidlaw}.

We can now define an ``exchange operator'' that smoothly carries the final
particle configuration around a closed path in ${\cal C}_n$.  Although this
exchange does not disturb the positions of the particles, it mixes up the
different sectors that contribute to the path integral.  Since these sectors
are weighted differently, in general, the exchange need not preserve the
amplitude.  This means that the amplitude need not be a single-valued function
of the $n$ positions of the final particles.  The effect of the exchange can be
expressed as the action of a linear operator acting on the amplitude, and
because the total probability sums to one, this operator is unitary.  By
considering the effect of two exchanges performed in succession, we readily see
that the exchange operators provide a unitary representation of the the group
$\pi_1({\cal C}_n)$.  Thus, a unitary representation of $\pi_1({\cal C}_n)$
acting on amplitudes (or wave functions) is a general feature of the quantum
mechanics of $n$ indistinguishable particles.  (The weight factors appearing in
the path integral also transform as a unitary representation of $\pi_1({\cal
C}_n)$.)

If the manifold is $R^d$ for $d\ge 3$, then $\pi_1({\cal C}_n)=S_n$, and the
exchange operators provide a unitary representation of the permutation group
$S_n$.  In addition to the familiar one-dimensional representations associated
with Bose and Fermi statistics, non-abelian representations
(``parastatistics'') are also possible in principle.  But it is known that, in
a local quantum field theory, parastatistics can always be reduced to Bose or
Fermi statistics by introducing additional degrees of freedom and a suitable
global symmetry that acts on these degrees of freedom \cite{doplicher}.
For $d=1$, in this
framework, no exchange is possible---the particles cannot pass through each
other---and there is no quantum statistics to discuss.

The case $d=2$ is the most interesting.  Then $\pi_1({\cal C}_n)$ is $B_n$, the
braid group on $n$ strands.  This is an infinite group with $n-1$ generators
$\sigma_1,\sigma_2,\cdots,\sigma_{n-1}$, where $\sigma_j$ may be interpreted as
a (counterclockwise) exchange of the particles in positions $j$ and $j+1$.
These generators obey the defining relations
\begin{equation}
\sigma_j\sigma_k=\sigma_k\sigma_j\; ,\quad |j-k|\ge 2\; ,
\end{equation}
and
\begin{equation}
\sigma_j\sigma_{j+1}\sigma_j=\sigma_{j+1}\sigma_j\sigma_{j+1}\; ,\quad
j=1,2,\cdots,n-2
\end{equation}
(The Yang-Baxter relation).  It follows from the Yang-Baxter relation that, in
a one-dimensional unitary representation of the braid group, all of the
$\sigma_j$'s are represented by a common phase $e^{i\theta}$.  This is anyon
statistics.  But non-abelian representations of the braid group may also arise
in local quantum field theories.  Indistinguishable particles in two dimensions
that transform under exchange as a non-abelian unitary representation of the
braid group are said to obey non-abelian statistics.

Our discussion of non-abelian vortices fits into the general framework outlined
above, but with an important caveat.  If the vortex flux takes values in an
unbroken local symmetry group $H(x_0)$, we treat two vortices with flux $a$ and
$b$ as ``indistinguishable'' if $b=hah^{-1}$ for some $h\in H(x_0)$, and if
both vortices have the same charge (transform as the same irreducible
representation of the centralizer $N(a)\cong N(b))$.  The philosophy is that
the
particles are regarded as indistinguishable if an exchange of the particles can
conceivably occur (in the presence of other particles with suitable quantum
numbers) without changing the quantum numbers assigned to the many-particle
configuration.  The caveat is that these ``indistinguishable'' particles are
not really identical.  For example, two vortices with flux $a$ and $b$ are
distinct---{\it e.g.}, the $a$ vortex will not annihilate the antiparticle of
the $b$ vortex---if $a\ne b$, even if $a$ and $b$ are in the same conjugacy
class.

This classification of the different types of ``indistinguishable'' vortices
can also be described in terms of the representation theory of a
quasi-triangular Hopf algebra, or ``quantum double''
\cite{bais92,orbifold,bantay}.  The quantum double
$D(H)$ associated with a finite group $H$ is an algebra that is
generated by global
gauge transformations and projection operators that pick out a particular value
of the flux.  A basis for the algebra is\footnote{In Ref.\
\cite{bais92,orbifold}, the notation ${\setbox0=\hbox{$\scriptstyle{h}$}
\hskip\wd0{\mathop{{\hbox{\vrule height 2.5ex depth 0pt \vrule width 2.5ex
height .4pt
 depth 0pt}}\llap{\vbox to 2.5ex{ \vfil
\hbox{$\scriptstyle{\box0}$\hskip 2.8ex} \vfil}}}
\limits_{a} }}$
is used for $P_h a$.}
\begin{equation}
\{P_h a\;, \quad h\,,a\in H\}\; ,
\end{equation}
where $P_h$ projects out the flux value $h$, and $a$ is a gauge transformation.
 Since the projection operators satisfy the relations
\begin{equation}
P_hP_g=\delta_{h,g}P_h\; ,\quad aP_h a^{-1}=P_{aha^{-1}}\; ,
\end{equation}
the multiplication law for the algebra can be expressed as
\begin{equation}
\left(P_h a\right)\cdot\left(P_g b\right)= \delta_{h, aga^{-1}}
\left(P_h ab\right)\; .
\end{equation}
An irreducible representation of the quantum double $D(H)$ can be
labeled $([a],R^{(a)})$, where $[a]$ denotes the conjugacy class that
contains $a\in H$ and $R^{(a)}$ is an irreducible representation of the
centralizer
$N(a)$ of $a$. This is the induced representation of $D(H)$ generated by the
representation $R^{(a)}$ of $N(a)$.  The space on which this
representation acts is a space of charged vortex states that transform
irreducibly under the global gauge transformations.
In order for an exchange contribution to an amplitude to interfere with the
direct amplitude, the two vortices being exchanged must belong to the
same irreducible representation of the quantum double.

(If a Chern-Simons term is added to the action of the underlying gauge
theory, the situation becomes somewhat more complicated \cite{bais92}. The
Chern-Simons term distorts the charge spectrum of vortices with a
specified value of the flux, and unremovable phases
can enter the multiplication law of the quantum double
\cite{bais92,orbifold,witdij}. The vortex
states may then transform as a projective (ray) representation under
gauge transformations.)

Consider a state of $n$ ``indistinguishable'' vortices, all with flux conjugate
to $a$, and all transforming as the representation $R^{(a)}$ of the centralizer
$N(a)$ (in other words, all of the vortices belong to the irreducible
representation $([a],R^{(a)})$ of the quantum double).
A basis for these states can be constructed, in which, at
each vortex
position, we assign a definite flux, and a definite basis state in the vector
space on which the representation $R^{(a)}$ acts.  Under exchange, these states
transform as a representation of $B_n$ that is in general non-abelian and
reducible.  This reducible representation can be decomposed into irreducible
components.  Each irreducible component describes an $n$-particle state obeying
definite ``braid statistics.''

The point that we wish to emphasize is that the exchange operator will
typically modify the quantum numbers that are assigned to the $n$ particle
positions.  Thus, physical observables, such as transition probabilities or
cross sections, need not be invariant under exchange.  Instead, the exchange
relates the value of the observable for one assignment of quantum numbers to
the particle positions to the value of the observable for another choice of
quantum numbers.
Correspondingly, as we stressed above, the observables are not single-valued
functions of the particle positions.  Only a subgroup of the braid group
returns the quantum numbers to their original values, and so preserves the
values of the physical observables.  (It is possible to restore the
single-valuedness of the many-body wave functions by introducing on the
configuration space a suitable connection with nontrivial holonomy.  The
existence of such a connection does not alter the essential physical point,
which is that ``indistinguishable'' vortices may have distinct quantum numbers
that can really be measured by an observer.)

Even distinguishable vortices have non-trivial Aharonov-Bohm interactions, so
it is appropriate to broaden this framework slightly.  We may consider a
many-particle state containing $n_1$ particles of type $1$ (with the type
characterized by the class of the flux $a$, and the charge $R^{(a)}$,
or, in other words, by the irreducible representation $([a],R^{(a)})$ of
the quantum double), $n_2$
particles of type 2, and so on.  Then an exchange of two particles is permitted
only if the particles are of the same type, and the wave function transforms as
a unitary representation of the ``partially colored braid group''
$B_{n_1,n_2\cdots}$ \cite{imbo,imbo91}.

Within this framework, a general connection between spin and statistics can be
derived, assuming the existence of an antiparticle corresponding to each
particle \cite{balastat,frolich,frolichetc}.  The essence of the connection is
that, if two particles are truly
{\it identical} (carry {\it exactly} the same quantum numbers), then an
exchange of the two particles can be smoothly deformed to a process in which no
exchange occurs, but one of the particles rotates by $2\pi$ \cite{balastat}.
(The reason that
the quantum numbers must be the same is that, for the deformation to be
possible, it is necessary for the antiparticle of the first particle to be able
to annihilate the second particle.)  It follows from the connection between
spin and statistics that the effect of an exchange of two objects that are
truly identical must be to modify the many-body wave function by the phase
$e^{2\pi i J}$, where $J$ is the spin of the object.  We have already remarked
in Sec.~\ref{sec:flux-charge} that this is true for non-abelian vortices with
the same flux and charge.  Thus, we find that non-abelian statistics is
perfectly compatible with the connection between spin and statistics.

There are deep connections between the theory of indistinguishable particles in
two spatial dimensions and conformally invariant quantum field theory in
two-dimensional {\it spacetime}.  These connections have been explored most
explicitly in the case of $(2+1)$-dimensional topological Chern-Simons
theories \cite{witten}, but appear to be more general \cite{frolichetc}.
There is a close mathematical analogy
between the {\it particle} statistics in two spatial dimensions that we have
outlined here, and the {\it field} statistics in two-dimensional conformal
field theory.  In the latter case, all correlation functions can be constructed
by assembling ``conformal blocks,'' and the conformal blocks typically
transform as a non-trivial unitary representation of the braid group when the
arguments of the correlation function are exchanged.  (See Ref.
\cite{chern} for a review.)  However, in discussions
of conformal field theory, it is usually the case that observables of interest
(the correlation functions themselves) are invariant under exchange.

\section{VORTEX-VORTEX SCATTERING}
\label{sec:vor-vor}

The holonomy interaction between vortices induces Aharonov-Bohm vortex-vortex
scattering, as pointed out by Wilczek and Wu \cite{wilczekwu} and Bucher
\cite{bucher}.  Suppose that a vortex
that initially carries flux $b$ is incident on a fixed vortex that initially
carries flux $a$.  Let us suppose, for now, that the vortices are uncharged.

To understand the behavior of the $b$ vortex propagating on the background of
the fixed $a$ vortex, it is convenient to adopt a path integral viewpoint.
Consider the two possible paths shown in Fig.~\ref{figD}.  If the vortex
follows the path
that passes below the scattering center, it will arrive at its destination with
flux $b$.  But if it follows the path that passes above the scattering center,
it arrives carrying the flux $aba^{-1}$.  Thus, if the flux of the scattering
center and the flux of the projectile do not commute, the contribution to the
path integral from paths that pass below does not interfere with the
contribution from paths that pass above.  Therefore, a plane wave propagating
on the background of the fixed vortex does not remain a plane wave---there is
nontrivial scattering.

More generally, the paths can be classified according to how many times they
wind around the scattering center (relative to some standard path).  The flux
of a $b$ vortex that winds around an $a$ vortex $k$ times is modified according
to
\begin{equation}
\label{kwind}
|b\rangle\to |(ab)^k b(ab)^{-k}\rangle\equiv |k\rangle \; .
\end{equation}
Since the unbroken gauge group $H$ is assumed to be finite, the flux eventually
returns to its original value, say after $n$ windings.

The flux of the scattered vortex, then, can take any one of $n$ values.  The
amplitude for
the vortex to arrive at the detector in the flux state $|k\rangle$ defined in
Eq.\ (\ref{kwind}) can be found by summing over all paths with winding number
congruent to $k$ modulo $n$.  Since only every $nth$ winding sector is included
in the amplitude $\psi_k$ for flux channel $k$, this amplitude is not a
periodic function of the polar angle $\phi$ with period $2\pi$; rather, the
period is $2\pi n$.  The $n$ amplitudes are related by the nontrivial monodromy
property
\begin{equation}
\label{kmono}
\psi_k(r,\phi+2\pi)=\psi_{k+1}(r,\phi) \;
\end{equation}
(where $\psi_{k+n}(r,\phi)\equiv\psi_k(r,\phi)$.)
Similarly, the exclusive cross section for flux channel $k$ is also
multi-valued:
\begin{equation}
\sigma_k(\theta-2\pi)=\sigma_{k+1}(\theta)\; ,
\end{equation}
where $\theta=\pi-\phi$ is the scattering angle.  The inclusive cross section
\begin{equation}
\sigma_{\rm inc}(\theta)=\sum_{k=0}^{n-1}\sigma_k(\theta)
\end{equation}
is single-valued.

As we stressed in the introduction, the multi-valuedness of the exclusive cross
sections is natural and unavoidable in this context.  Whenever we assign a flux
to a non-abelian vortex, we are implicitly adopting a conventional procedure
for measuring the flux.  For example, the procedure might be to carry the
vortex to the ``vortex bureau of standards'' and analyze it there by performing
Aharonov-Bohm interference experiments with various charged particles.  Then
the multi-valuedness arises because, if we carry a vortex in the flux state
$|k\rangle$ once around the scattering center (counterclockwise) before
returning it to the bureau of standards, the analysis will identify it as the
flux state $|k+1\rangle$.

For each value of the scattering angle, we might choose a standard path along
which the vortex is to be returned to the bureau for analysis after the
scattering event.  For example, we might decide to carry it home through the
upper half plane for $\theta\in [0,\pi)$ and through the lower half plane for
$\theta\in [-\pi,0)$, as shown in Fig.~\ref{figE}.  Then the exclusive cross
sections are
single-valued, but are discontinuous at $\theta=0$:
\begin{equation}
\sigma_k(\theta=0^+)=\sigma_{k+1}(\theta=0^-)\; .
\end{equation}
The choice of a standard path amounts to an arbitrary restriction of the
$n$-valued exclusive cross sections to a single branch.

In a sense, the multi-valuedness of the wave functions, and of the exclusive
cross sections, arises because we have insisted on expressing the flux of the
vortices in terms of a multi-valued basis---that basis defined by parallel
transport of the flux in the background gauge potential of the scattering
center.  The propagation of the projectile on this background is really
non-singular, and the multi-valuedness of the amplitudes actually compensates
for the multi-valuedness of the basis.  This is quite analogous to the
``singular-gauge'' description of ordinary abelian Aharonov-Bohm scattering.
There, expressing the phase of the electron wave function relative to a basis
defined by parallel transport is equivalent to performing a singular gauge
transformation that gauges away the vector potential and introduces a
discontinuity in the wave function.  The difference in the non-abelian case is
that the discontinuity corresponds to a jump in observable quantum numbers of
the projectile, as explained above.  It is natural to use the multi-valued
basis because it reflects what a team of experimenters would really find if
they brought their detectors together to calibrate them alike.

Mathematically, finding the Aharonov-Bohm amplitude for a vortex propagating on
the background of a fixed vortex is equivalent to finding the amplitude for a
free particle propagating on an $n$-sheeted surface.  (The closely related
problem of a free particle propagating on a cone has been discussed in
connection with $2+1$ dimensional general relativity \cite{hooft,deser}.)  The
most convenient way
to solve the problem is to transform to a basis of ``monodromy eigenstates,''
since for the elements of this basis the scattering reduces to abelian
Aharonov-Bohm scattering.  If the $\psi_k$'s obey the monodromy property Eq.\
(\ref{kmono}), then the monodromy eigenstate basis is
\begin{equation}
\chi_l={1\over\sqrt{n}}\sum_{k=0}^{n-1}e^{-2\pi i kl/n}\psi_k\; ,
\end{equation}
with the property
\begin{equation}
\chi_l(r,\phi+2\pi)=e^{2\pi i l/n}\chi_l(r,\phi)\; .
\label{monoeigen}
\end{equation}
These monodromy eigenstates correspond to states of the two-vortex system that
have definite charge, in the sense that they are eigenstates of the gauge
transformation $ab\in H$, where $ab$ is the total flux.

We may think of the wave functions $\chi_l$ as the coefficients in an expansion
of a single-valued wave function in a multi-valued basis.  That is, we can
express a single-valued wave function as
\begin{equation}
|\psi\rangle=\sum_{r,\phi,l} |r,\phi, l\rangle\langle r, \phi, l |\psi\rangle\;
,
\end{equation}
where the basis $|r,\phi,l\rangle$ is ``twisted'' according to
\begin{equation}
|r,\phi+2\pi,l\rangle =e^{-2\pi il/n}|r,\phi,l\rangle\; .
\label{basismono}
\end{equation}
The coefficients $\chi_l(r,\phi)=\langle r,\phi,l|\psi\rangle$ inherit the
property Eq.~(\ref{monoeigen}) from the property Eq.~(\ref{basismono}) of the
basis.

By standard methods \cite{aharonov}, we can find the solution to the
free-particle
nonrelativistic Schr\"odinger equation that obeys the condition
\begin{equation}
\chi_\alpha(\phi+2\pi)=e^{2\pi i\alpha}\chi_\alpha(\phi)\; ,
\quad 0\le \alpha < 1\; ,
\label{alphadef}
\end{equation}
and matches a plane wave incoming from $\phi=0$.  The asymptotic large-$r$
behavior of this solution is
\begin{equation}
\chi_\alpha\sim e^{-i\vec{p}\cdot \vec{x}}
+{e^{i p r}\over\sqrt{r}}f_\alpha(\phi)\; ,\quad -\pi<\phi<\pi\; ,
\label{defineamp}
\end{equation}
where
\begin{equation}
f_\alpha(\phi)={e^{-i\pi/4}\over\sqrt{2\pi p}}
\left({1\over
1+e^{i\phi}}\right)e^{i\alpha\phi}\left(e^{-i\alpha\pi}-e^{i\alpha\pi}\right)
\; , \quad 0\le \alpha < 1\; .
\label{eq:ABamp}
\end{equation}
Here $e^{-i\alpha\pi}$ is the phase shift for the partial waves with
non-negative integer part of the orbital angular momentum and $e^{i\alpha\pi}$
is the phase shift for the partial waves with negative integer part of the
orbital angular momentum.  The semiclassical interpretation is that wave
packets that pass above and below the scattering center acquire a relative
phase $e^{2\pi i\alpha}$, the Aharonov-Bohm phase.

There are two subtleties concerning Eq.~(\ref{defineamp}) and (\ref{eq:ABamp})
that deserve comment.
The first subtlety (which is not very important for what follows), is that
there is an order of limits ambiguity in the evaluation of the amplitude---the
limit $r\to\infty$ does not commute with the limit $\phi\to \pm\pi$
\cite{hagen}.  In
Eq.~(\ref{defineamp}) and (\ref{eq:ABamp}),
we have taken $r\to\infty$ for fixed $\phi$ between $-\pi$ and $\pi$.  Thus,
$\chi_\alpha$
actually satisfies Eq.~(\ref{alphadef}), although the first term in
the asymptotic form Eq.~(\ref{defineamp}) appears not to.  (For large $r$, the
phase of the
plane wave in Eq.~(\ref{alphadef}) suddenly advances by $e^{2\pi
i\alpha}$ as $\phi$ increases through a narrow wedge near $\phi=\pi$.  Of
course, if we construct localized wave packets, then the unscattered wave has
support at $\phi=0,\pm\pi$ as $r\to\infty$, and the form of the plane wave away
from the forward direction is of no consequence anyway.)
The second subtlety, which is very important for what follows, concerns the
$\alpha$ dependence of the amplitude.  The monodromy condition
Eq.~\ref{alphadef} depends only on $\alpha-[\alpha]$, where $[\alpha]$ denotes
the greatest integer less than or equal to $\alpha$.  Thus, as one can
explicitly verify, the amplitude $f_{\alpha}(\phi)$, when $\alpha$ is not
restricted to lie in the range $[0,1)$, takes the same form as
Eq.~(\ref{eq:ABamp}), but with $\alpha$ replaced by $\alpha-[\alpha]$.  The
somewhat surprising feature is that, as a function of $\alpha$,
$f_\alpha(\phi)$ is not differentiable when $\alpha$ is an integer.

The form Eq.~(\ref{eq:ABamp}) for the scattering amplitude in the monodromy
eigenstate basis is
readily generalized to an arbitrary basis, if we express it in terms of the
braid operator ${\cal R}$, the square root of the monodromy operator ${\cal
M}$.  The general monodromy condition satisfied by the wave function can be
expressed as
\begin{equation}
\psi(\phi + 2\pi)={\cal M}\psi(\phi)\; ,
\end{equation}
where ${\cal M}$ is a unitary matrix acting on internal indices.  Then the
basis-independent form for the scattering amplitude is
\begin{equation}
\label{ABamp}
\langle{\rm out}|f(\phi)|{\rm in}\rangle
={e^{-i\pi/4}\over\sqrt{2\pi p}}
\left({1\over 1+e^{i\phi}}\right)
\langle{\rm out}|{\cal R}^{\phi/\pi}\left({\cal R}^{-1}-
{\cal R}\right)|{\rm in}\rangle
\; ,
\end{equation}
where ${\cal R}$ is defined by ${\cal R}^2={\cal M}$, and $|{\rm in}\rangle$,
$|{\rm out}\rangle$ denote the incoming and outgoing wave functions in internal
space.  This definition of ${\cal R}$ leaves an ambiguity in ${\cal
R}^{(\phi/\pi \mp 1)}$, and it is important to resolve this ambiguity
correctly.
Acting on an eigenstate of ${\cal M}$ with
\begin{equation}
{\cal M}=e^{2\pi i \alpha}\; ,
\end{equation}
we define
\begin{equation}
{\cal R}^{(\phi/\pi \mp 1)}\equiv e^{i(\alpha-[\alpha])(\phi\mp \pi)}\; .
\label{braidpowerdef}
\end{equation}
In Eq.~(\ref{ABamp}), the state $|{\rm in\rangle}$ is expressed in terms of an
arbitrary basis, and we have assumed that the state $|{\rm out}\rangle$ is
expressed in terms of a basis that is obtained by parallel transport of the
in-basis.  This out-basis is
multi-valued, so we have in effect evaluated the amplitude in a ``singular
gauge.''

{}From Eq.~(\ref{ABamp}), we obtain the cross section
\begin{equation}
\label{ABsigma}
\sigma_{{\rm in}\to {\rm out}}(\phi)=\left|f(\phi)\right|^2
={1\over 2\pi p}\left({1\over 4\cos^2{\phi/2}}\right)
\left|\langle{\rm out}|{\cal R}^{\phi/\pi}\left({\cal R}^{-1}-
{\cal R}\right)|{\rm in}\rangle\right|^2\; .
\end{equation}
By summing $|\rm{out}\rangle$ over a complete basis, we obtain the inclusive
cross
section
\begin{equation}
\label{sigmaincl}
\sigma_{{\rm in}\to {\rm all}}(\theta)=
{1\over 2\pi p}\left({1\over \sin^2{\theta/2}}\right)
{1\over 2}\left(1-{\rm Re}\langle{\rm in}|{\cal R}^2|{\rm in}\rangle\right)\; ,
\end{equation}
where $\theta=\pi-\phi$ is the scattering angle; this is the formula derived by
Verlinde \cite{verlinde}.

For monodromy eigenstates with ${\cal M}=e^{2\pi i\alpha}$, Eq.\
(\ref{ABsigma}) reduces to the familiar form of the Aharonov-Bohm cross
section,
\begin{equation}
\sigma_\alpha(\theta)={1\over 2\pi p}\left({
\sin^2{\pi \alpha}\over \sin^2{\theta/2}}\right)\; ,
\end{equation}
which is a single-valued function of the scattering angle.  But the recurring
theme of this paper is that it is often convenient to express the scattering
states in terms of a basis other than the monodromy eigenstate basis. Then the
exclusive cross sections are in general multi-valued, but the inclusive cross
section (summed over all possible final state quantum numbers) is always
single-valued.

Returning to the special case of (uncharged) vortex-vortex scattering, we
obtain the amplitude in the flux eigenstate basis by coherently summing the
monodromy eigenstate amplitudes with appropriate phases,
\begin{eqnarray}
&&\langle k|f(\phi)|k=0\rangle={1\over n}\sum_{l=0}^{n-1}e^{2\pi i kl/n}
f_{l/n}(\phi) \nonumber\\
&=&{e^{-i\pi/4}\over\sqrt{2\pi p}}\left({i\over 2n}\right){\sin(\pi/n)\over
\sin\left[{1\over 2n}\left(\phi+(2k+1)\pi\right)\right]
\sin\left[{1\over 2n}\left(\phi+(2k-1)\pi\right)\right]}\; .
\label{fixedABamp}
\end{eqnarray}
This formula has the expected monodromy property
\begin{equation}
\langle k|f(\phi+2\pi)|k=0\rangle=\langle k+1|f(\phi)|k=0\rangle\; .
\end{equation}
(Eq.\ (\ref{fixedABamp}) is actually a special case of the the formula derived
in (2+1)-dimensional gravity by 't Hooft \cite{hooft} and Deser and Jackiw
\cite{deser}.)

This amplitude has the infinite forward peak that is characteristic of
Aharonov-Bohm scattering.  For $\phi=\pi$, the infinite peak occurs in the flux
channels $k=0,-1$ and for $\phi=-\pi$, it occurs in the channels $k=1,0$.  For
$\phi$ near $\pi$, the leading behavior of the amplitude is
\begin{equation}
\langle k=0|f(\phi)|k=0\rangle\sim -\langle k=-1|f(\phi)|k=0\rangle
\sim {e^{-i\pi/4}\over\sqrt{2\pi  p}}\left({i\over\phi-\pi}\right)\; .
\end{equation}
This leading behavior has a simple interpretation.  From a path integral
viewpoint, the forward peak is generated by paths that pass above or below the
scattering center with a large impact parameter, without any winding around the
center.  If the projectile passes above, it is detected near $\phi=\pi$ as a
$k=0$ vortex (or near $\phi=-\pi$ as a $k=1$ vortex); if it passes below, it is
detected near $\phi=\pi$ as a $k=-1$ vortex (or near $\phi=-\pi$ as a $k=0$
vortex).  Near $\phi=\pi$, the amplitude in the $k=0,-1$ channels is equivalent
to the diffraction pattern generated by a ``sharp edge,'' since paths that wind
$n$ times around the scattering center make a negligible contribution.  The
near-forward amplitude in the $k=0$ channel comes from summing all of the
partial waves with non-negative angular momentum, and the near-forward
amplitude in the $k=-1$ channel comes from summing the partial waves with
negative angular momentum.  Thus, the forward peak in each channel is half as
strong as the forward peak for ``maximal'' ($\alpha=1/2$) abelian Aharonov-Bohm
scattering.

The inclusive cross section (obtained by summing over all possible final flux
channels) can be immediately read off from Eq.\ (\ref{sigmaincl}).  If the
projectile is a flux eigenstate, and the scattering center is a flux eigenstate
whose flux does not commute with that of the projectile, then we have $\langle
{\rm in}|{\cal R}^2|{\rm in}\rangle=0$, and the inclusive cross section takes
the universal form
\begin{equation}
\sigma_{{\rm flux~eigenstate}\to {\rm all}}(\theta)=
\left({1\over 2}\right){1\over 2\pi p}\left({1\over \sin^2{\theta/2}}\right)\;
;
\end{equation}
that is, half the cross section for maximal Aharonov-Bohm scattering.

So far, we have assumed that the vortex that is being scattered carries no
charge.  Let us briefly comment on how the analysis is modified when the
scattered vortex is charged.

Suppose that the vortex with flux $a$ transforms as some irreducible
representation $D^{R^{(a)}}$ of $N(a)$, and that the vortex with flux $b$
transforms as some irreducible representation $D^{R^{(b)}}$ of $N(b)$.  And
suppose as before that the fluxes return to their original values after the
monodromy operator acts $n$ times (that is, after the $b$ vortex winds around
the $a$ vortex $n$ times).  For charged vortex states, although ${\cal M}^n$
preserves the flux values, it acts on the vortex pair as a nontrivial $N(a)
\otimes N(b)$ transformation.  Specifically, we have
\begin{equation}
{\cal M}^n:|a\rangle\otimes |b\rangle\to
D^{R{(a)}}[(ab)^na^{-n}]|a\rangle\otimes D^{R{(b)}}[(ab)^n b^{-n}]|b\rangle\; .
\end{equation}
Note that, since by assumption $(ab)^n a(ab)^{-n}=a$ and $(ab)^n b (ab)^{-n}=b$
(because ${\cal M}^n$ preserves the fluxes), $(ab)^na^{-n}\in N(a)$ and $(ab)^n
b^{-n}\in N(b)$.

For the case of scattering a $b$ vortex off of a fixed $a$ vortex, we consider
the states $|k\rangle$ defined by
\begin{equation}
|k\rangle\equiv {\cal M}^k|b\rangle\; , \quad k=0,1,2,\dots,n-1\;,
\end{equation}
with
\begin{equation}
{\cal M}^n|k=0\rangle=D^{R^{(b)}}[(ab)^nb^{-n}]|k=0\rangle\; .
\end{equation}
To diagonalize the monodromy operator, we first diagonalize the unitary
transformation $D^{R^{(b)}}[(ab)^nb^{-n}]$.  Corresponding to each eigenstate
of this operator with eigenvalue $e^{2\pi i \beta}$ are a set of monodromy
eigenstate wave functions
\begin{equation}
\chi_{l,\beta}={1\over \sqrt{n}}\sum_{k=0}^{n-1}e^{-2\pi i k
(l+\beta)/n}\psi_{k,\beta}\; ,
\end{equation}
with the property
\begin{equation}
\chi_{l,\beta}(r,\phi+2\pi)=e^{2\pi i (l+\beta)/n}\chi_{l,\beta}(r,\phi)\; .
\end{equation}
For particular charged states with specified flux, we may evaluate
Eq.~(\ref{ABamp}) by coherently superposing the Aharonov-Bohm amplitudes for
these monodromy eigenstates.

\section{INDISTINGUISHABLE VORTICES}

The effects of quantum statistics can be seen in the two-body scattering of
indistinguishable particles, because exchange scattering can occur; it is
possible to lose track of ``who's who.''  In the case of non-abelian vortices,
the exchange effects are more subtle than for abelian anyons---in general,
whether two vortices behave like identical or distinct particles when
they are brought together depends on their {\it history}.  Suppose that two
identical vortices each carry the flux $a\in H$.  If one of the vortices should
voyage around another vortex with flux $b$, and then return to its partner, it
would then carry flux $bab^{-1}$.  Hence, if $a$ and $b$ do not commute, it
would now  be distinct from the other $a$ vortex.

For exchange effects to occur in vortex-vortex scattering, the braid operator
must have an orbit of odd order acting on the two vortex state.  That is,
${\cal R}^n$ must preserve the two vortex state for some odd $n$.  If so, there
will be a contribution to the vortex-vortex scattering amplitude in which the
two vortices change places, that interferes with the direct amplitude.

As a simple example, consider the permutation group on three objects $S_3$,
where the fluxes are two distinct two-cycles.  Then the braid operator defined
by Eq.\ (\ref{braidop}) has the orbit
\begin{eqnarray}
{\cal R}:\quad |(12),(23)\rangle&\to &|(13),(12)\rangle\nonumber\\
&\to & |(23),(13)\rangle\nonumber\\
&\to & |(12),(23)\rangle\; ,
\end{eqnarray}
of order 3. (See Fig.~\ref{figF}.) Thus, there is an exchange contribution to
the scattering of a
(12) vortex and a (23) vortex.  (In this case, the centralizer of the total
flux is $Z_3$, and the braid eigenstates are the linear combinations of these
three states that have definite $Z_3$ charge.)

Two vortices whose flux belongs to the same conjugacy class of the unbroken
group $H$ have the same mass, and we can easily derive a formula for the
vortex-vortex scattering amplitude in the center of mass frame, using the same
methods as in the previous section.  This formula will incorporate the exchange
effects whenever the braid operator has an odd orbit acting on the two-vortex
state.  The two-body wave function in the center of mass frame will now have
the property
\begin{equation}
\psi(r,\phi+\pi)={\cal R}\psi(r,\phi)\; ,
\end{equation}
where the braid operator ${\cal R}$ is a unitary matrix acting on the internal
indices of the wave function.  The problem is to solve the free-particle
Shr\"odinger equation subject to this condition.

If the two-body state is a ``braid eigenstate,''
\begin{equation}
\label{anybraid}
\chi_\alpha(r,\phi+\pi)=e^{i\pi\alpha}\chi_\alpha(r,\phi)\; ,\quad
0\le \alpha < 2\; ,
\end{equation}
then the problem is equivalent to anyon-anyon scattering, with statistical
phase $e^{i\theta}=e^{i\pi \alpha}$.  We can find the
solution to the free-particle Schr\"odinger equation that obeys Eq.\
(\ref{anybraid}) and matches plane waves coming from $\phi=0$ and $\phi=\pi$.
The asymptotic large-r behavior of this solution is \cite{mrwilczek}
\begin{equation}
\chi_\alpha\sim \left(e^{-i\vec{p}\cdot \vec{x}}+
e^{i\alpha \pi}e^{i\vec{p}\cdot \vec{x}}\right)
+{e^{i p r}\over\sqrt{r}}f_\alpha(\phi)\; ,\quad 0<\phi<\pi\; ,
\label{braidampdef}
\end{equation}
where
\begin{equation}
f_\alpha(\phi)={e^{-i\pi/4}\over\sqrt{2\pi p}}
\left({2\over
1-e^{2i\phi}}\right)e^{i\alpha\phi}\left(e^{-i\alpha\pi}-e^{i\alpha\pi}\right)
\; , \quad 0\le\alpha< 2\; .
\label{eq:exchamp}
\end{equation}
(As in our discussion of scattering off a fixed target, we remark that the
limit $r\to\infty$ does not commute with the limit $\phi\to 0,\pi$
\cite{hagen}.  Thus, $\chi_\alpha$ actually satisfies Eq.~(\ref{anybraid}),
although the first term in the asymptotic form Eq.~(\ref{braidampdef}) appears
not to.)
In an arbitrary basis, in which the braid operator is not necessarily diagonal,
we have
\begin{equation}
\label{ABampexch}
\langle{\rm out}|f(\phi)|{\rm in}\rangle
={e^{-i\pi/4}\over\sqrt{2\pi p}}
\left({2\over 1-e^{2i\phi}}\right)
\langle{\rm out}|{\cal R}^{\phi/\pi}\left({\cal R}^{-1}-
{\cal R}\right)|{\rm in}\rangle
\; ,
\end{equation}
where $|{\rm in}\rangle$, $|{\rm out}\rangle$ denote the incoming and outgoing
two-body wave functions in internal space.  As in our discussion of scattering
off of a fixed center, there is an ambiguity in the evaluation of ${\cal
R}^{(\phi/\pi \mp 1)}$, and we must now resolve this ambiguity slightly
differently than before.  If $\alpha$ is not restricted to the range $[0,2)$,
then $\alpha$ must be replaced by $\alpha-[[\alpha]]$ in Eq.~\ref{eq:exchamp},
where $[[\alpha]]$ denotes the greatest {\it even} integer less than or equal
than $\alpha$.  Thus, acting on an eigenstate of ${\cal R}$ with eigenvalue
\begin{equation}
{\cal R} = e^{i\pi \alpha}\; ,
\end{equation}
we define ${\cal R}^{(\phi/\pi\mp 1)}$ by
\begin{equation}
{\cal R}^{(\phi/\pi \mp 1)}=e^{i(\alpha- [[\alpha]])(\phi\mp \pi)}\; .
\end{equation}
The cross section is
\begin{equation}
\label{ABsigmaexch}
\sigma_{{\rm in}\to {\rm out}}(\phi)=\left|f(\phi)\right|^2
={1\over 2\pi p}\left({1\over \sin^2{\phi}}\right)
\left|\langle{\rm out}|{\cal R}^{\phi/\pi}\left({\cal R}^{-1}-
{\cal R}\right)|{\rm in}\rangle\right|^2\; .
\end{equation}
By summing $|\rm{out}\rangle$ over a complete basis, we obtain the inclusive
cross section
\begin{equation}
\label{sigmainclexch}
\sigma_{{\rm in}\to {\rm all}}(\theta)=
{1\over 2\pi p}\left({1\over \sin^2{\theta}}\right)
2\left(1-{\rm Re}\langle{\rm in}|{\cal R}^2|{\rm in}\rangle\right)\; ,
\end{equation}
where $\theta=\pi-\phi$ is the scattering angle.

The general problem can be solved by expressing the two-body state as a linear
combination of braid eigenstates, and then coherently superposing the
anyon-anyon amplitudes.  In the case of (uncharged) vortex-vortex scattering,
if the initial state is a vortex with flux $a$ coming from $\phi=\pi$ and a
vortex with flux $b$ coming from $\phi=0$, then let us denote by $|k\rangle$
the state obtained when the braid operator ${\cal R}$ defined by Eq.\
(\ref{braidop}) acts on the initial state $k$ times
\begin{equation}
|k\rangle\equiv {\cal R}^k |a,b\rangle \; .
\end{equation}
Suppose that the two-vortex state returns to the initial state after ${\cal R}$
acts $n$ times.
(Note that, in a departure from the notation of the previous section, $k$ and
$n$ now denote the number of times the {\it braid} operator acts on the initial
state, rather than the monodromy operator ${\cal M}={\cal R}^2$.)
Then,
\begin{equation}
\chi_{2l/n}=\sum_{k=0}^{n-1}e^{-2\pi i k l/n} |k\rangle
\end{equation}
is a braid eigenstate with eigenvalue $e^{i\pi\alpha}=e^{2\pi il/n}$, and the
scattering amplitude in the flux eigenstate basis is
\begin{eqnarray}
&&\langle k|f(\phi)|k=0\rangle={1\over n}\sum_{l=0}^{n-1}e^{2\pi i kl/n}
f_{2l/n}(\phi) \nonumber\\
&=&{e^{-i\pi/4}\over\sqrt{2\pi p}}\left({i\over n}\right){\sin(2\pi/n)\over
\sin\left[{1\over n}\left(\phi+(k+1)\pi\right)\right]
\sin\left[{1\over n}\left(\phi+(k-1)\pi\right)\right]}\; .
\label{cmscattform}
\end{eqnarray}
This formula has the desired property
\begin{equation}
\langle k|f(\phi+\pi)|k=0\rangle=\langle k+1|f(\phi)|k=0\rangle\; .
\label{exscattform}
\end{equation}
Eq.\ (\ref{cmscattform}) applies for any value of $n$, but there is an exchange
contribution to the amplitude only for odd $n$.  (Note that, if $n$ and $k$ are
even, Eq.~(\ref{exscattform}) precisely coincides with Eq.~(\ref{cmscattform}),
as one would expect.)

The amplitude has the expected infinite peak at $\phi=\pi$ in the channels
$k=0,-2$ and at $\phi=0$ in the channels $k=\pm 1$.  As in our discussion of
scattering off of a fixed center, these peaks are generated by paths in which
the two vortices pass one another with a large impact parameter, without any
winding.  If the vortex incident from the right passes above the vortex
incident from the left, then, with our conventions, a $k=0$ state is detected
near $\phi=\pi$, and a $k=1$ state is detected near $\phi=0$.  If the vortex
incident from the right passes below, then a $k=-2$ state is detected near
$\phi=\pi$, and a $k=-1$ state is detected near $\phi=0$.

\section{CONTINUOUS SYMMETRY: THE ALICE VORTEX}
\label{sec:alice}

So far, we have assumed that the unbroken local symmetry group is a discrete
group.  In this section, we will briefly consider the properties of non-abelian
vortices when the gauge group is continuous.

If the unbroken gauge group has a non-abelian Lie algebra, then the gauge
interaction is presumably confining. In fact, even if the Lie algebra is
abelian (a product of $U(1)$'s), then charge is logarithmically confined in two
spatial dimensions.  That is, the Coulomb energy of a charged object is
logarithmically infrared divergent.  Nevertheless, we might be interested in
the Aharonov-Bohm interactions of vortices and charged particles on distance
scales that are small compared to the confinement scale, or under circumstances
where the Coulomb energy can be safely neglected.

Strictly speaking, there is no Aharonov-Bohm amplitude for the scattering
of a charged particle off of a vortex, because  there are no asymptotic charged
states.  Still, the formalism discussed
in this paper finds some application.  We can imagine placing a compensating
charge far away from the scattering center, and consider the scattering of a
wave packet in a bounded region that is small compared to the distance to the
compensating charge (or small compared to the confinement distance scale).
Furthermore, the charge of a particle behaves like $\hbar
e$, where $e$ is a (classical) gauge coupling, so Coulomb effects are of order
$(\hbar e)^2$, and are higher order corrections to Aharonov-Bohm scattering in
the semiclassical (small $\hbar$) limit.  Under suitable conditions, the
deflection of the wave packet is described to good accuracy by our general
formula for the Aharonov-Bohm amplitude, Eq.~(\ref{ABamp}).

The case of vortex-vortex scattering is more complicated.  We can imagine
scattering two vortices that are flux eigenstates.  (More properly, in the case
of continuous gauge symmetry, we should consider narrow ``flux wave packets,''
superpositions of flux eigenstates with small dispersion.)  However, a pair of
flux eigenstates does not have definite charge;  when the state of the pair is
decomposed into charge eigenstates, the states of nonzero charge have infrared
divergent Coulomb energy.  Again, there is a need for a compensating charge.
But in this case, the value of the compensating charge must be correlated with
the state of the vortex pair.  If we trace over the state of the compensating
charge, we obtain a density matrix for the vortex pair that is an {\it
incoherent} superposition of charge eigenstates.  Thus, the ``scattering cross
section'' is an incoherent sum of the cross sections for the various charge (or
braid) eigenstates,  and Eq.~(\ref{ABamp}) does not apply.

To make the discussion more definite, let us consider the simplest model that
exhibits these features, the ``Alice'' model
\cite{schwarz,alfordbenson,preskillkrauss,bucherlopreskill}.  The unbroken
symmetry group in
this case is the semi-direct product of $U(1)$ with $Z_2$.  The group has a
component connected to the identity, the $U(1)$ subgroup, that can be
parametrized as
\begin{equation}
\{e^{i\omega Q}\; ,\quad 0\le\omega <2\pi\}\; ,
\end{equation}
where $Q=\sigma_3$ is the $U(1)$ generator.  There is also a component that is
not connected to the identity,
\begin{equation}
\{i\sigma_2 e^{i\omega Q}\; , \quad 0\le \omega <2\pi\}\; .
\end{equation}
Each element of the disconnected component anticommutes with $Q$.  Thus, the
Alice model can be characterized as a generalization of electrodynamics in
which charge conjugation is a {\it local} symmetry.

An ``Alice vortex'' carries flux that takes a value in the disconnected
component of this group.  The monodromy operator associated with transport
around this vortex, acting on the defining representation of the group, is
\begin{equation}
{\cal M}(\omega) = e^{-i\omega Q/2}i\sigma_2 e^{i\omega Q/2}\; .
\label{alicemono}
\end{equation}
Because ${\cal M}$ anticommutes with $Q$, when a charged particle is
transported around the vortex, its charge flips in sign.  This monodromy
property induces Aharonov-Bohm scattering of the charge eigenstates.  Using the
prescription Eq.~(\ref{braidpowerdef}), it is
straightforward to compute
\begin{equation}
{\cal R}^{\phi/\pi}\left({\cal R}^{-1} - {\cal R}\right)=
e^{-i\omega Q/2}\left(-i\sqrt{2}\right)e^{i\phi/2}
\pmatrix{\cos\phi/4&-\sin\phi/4\cr\sin\phi/4&\cos\phi/4\cr}e^{i\omega Q/2}\; .
\end{equation}
{}From Eq.~(\ref{ABamp}), we thus obtain the cross section for scattering of
charge eigenstates off of a fixed Alice vortex,
\begin{equation}
\sigma_{\pm}(\theta)={1\over 2\pi p}{1\pm \sin\theta/2\over 4\sin^2\theta/2}\;
;
\label{alicecross}
\end{equation}
here, $\sigma_+$ denotes the cross section when the scattered charge has the
same sign as the original projectile, and $\sigma_-$ is the cross section for
charge-flip scattering.  Note that these exclusive cross sections respect the
relation Eq.~(\ref{eq:alicecrossmono}) anticipated in the introduction.

The case of a charged particle scattering from an Alice vortex is quite similar
to the case of vortex-vortex scattering considered in Sec.~\ref{sec:vor-vor},
where the orbit of the monodromy operator has order $n=2$.  There is an
important difference, however---the monodromy operator Eq.~(\ref{alicemono})
squares to $-1$ rather than $1$.  The property ${\cal M}^2=-1$ holds whenever
the charge of the projectile is odd, and hence the cross section
Eq.~(\ref{alicecross}) applies for any odd charge.  The vanishing of $\sigma_-$
in the backward direction is easily seen to be a consequence of ${\cal
M}^2=-1$;  the trajectories with positive and negative {\it odd} winding number
interfere destructively at $\theta=\pi$.  If the charge of the projectile is
even, then ${\cal M}^2=1$, and the cross section is given by
Eq.~(\ref{fixedABamp}) for $n=2$, with $k=0$ corresponding to $\sigma_+$ and
$k=1$ to $\sigma_-$.

Now consider the case of vortex-vortex scattering, in the flux eigenstate
basis.  We denote by $|\omega\rangle$ the vortex state with flux
$i\sigma_2e^{i\omega Q}$.  According to Eq.~(\ref{braidop}), the effect of an
exchange on a state of two vortices, each with definite flux, can be expressed
as
\begin{equation}
{\cal R}:  |\omega_1,\omega_2\rangle \to |2\omega_1-\omega_2, \omega_1\rangle
\; .
\end{equation}
The exchange preserves the ``total flux'' $i\sigma_2e^{i\omega_1
Q}i\sigma_2e^{i\omega_2 Q}=e^{i(\omega_2-\omega_1)Q}\equiv e^{i\omega_{\rm
tot}Q}$, so an alternative notation is
\begin{equation}
{\cal R}: |\omega_1;\omega_{\rm tot}\rangle\to |\omega_1-\omega_{\rm
tot};\omega_{\rm tot}\rangle\; ,
\end{equation}
with the flux $\omega_2=\omega_{\rm tot}+\omega_1$ of the second vortex
suppressed.

The two vortex state can be decomposed into states with definite transformation
properties under the centralizer of the total flux, which is $U(1)$.  These
charge eigenstates also diagonalize the braid operator.  The action of $U(1)$
on the flux eigenstates is
\begin{equation}
e^{i\epsilon Q}: |\omega_1; \omega_{\rm tot}\rangle\to
|\omega_1-2\epsilon;\omega_{\rm tot}\rangle\; ,
\end{equation}
and the charge eigenstates are
\begin{equation}
|q,\omega_{\rm tot}\rangle={1\over \sqrt{\pi}}\int^{2\pi}_0 d\omega'
e^{iq\omega'}
|2\omega';\omega_{\rm tot}\rangle\; ,
\end{equation}
where the charge $q$ is an {\it even} integer.  The braid operator acts on the
charge
eigenstates according to
\begin{equation}
{\cal R}:|q,\omega_{\rm tot}\rangle\to e^{iq\omega_{\rm tot}}|q,\omega_{\rm
tot}\rangle\; .
\end{equation}
Formally, we can find the amplitude for a vortex with flux $\omega_1$ to
scatter from a fixed center with flux $\omega_2=\omega_{\rm tot}+\omega_1$ by
applying Eq.~(\ref{ABamp}).  The result is
\begin{eqnarray}
&\langle&\omega_1';\omega_{\rm tot}~{\rm out}|f(\phi)|\omega_1;\omega_{\rm
tot}~{\rm in}\rangle\\
&=&{e^{-i\pi/4}\over\sqrt{2\pi p}}\left({1\over 1+e^{i\phi}}\right)
{1\over \pi}\sum_q e^{iq(\omega'-\omega)}\left(e^{i(q\omega_{\rm
tot}-[q\omega_{\rm tot}])(\phi/\pi-1)}-e^{i(q\omega_{\rm tot}-[q\omega_{\rm
tot}])(\phi/\pi+1)}\right)\;
\end{eqnarray}
where $q$ is summed over even integers.  We note that it is essential to
subtract away the integer part of $q\omega_{\rm tot}$ in order to obtain the
correct result.  For
example, if $\omega_{\rm tot}$ is rational, then the amplitude has support only
for discrete values of $\omega'-\omega$.  This would not have worked if the
integer part had not been subtracted.

However, as noted above, this analysis is moot, because of the need to deal
with the infrared divergent Coulomb energy of the states with $q\ne 0$.  One
way to screen the charge is to place another vortex pair far away, such that
the four-vortex system carries total charge zero.  But however we arrange to
screen the charge, the state of the vortex pair we are studying will be
correlated with the state of the compensating charge (unless the vortex pair is
in a charge eigenstate).  For example, our flux eigenstate becomes
\begin{equation}
|\omega_1;\omega_{\rm tot}\rangle\to{1\over\sqrt{\pi}}\sum_q e^{-iq\omega_1}
|q;\omega_{\rm tot}\rangle\otimes|-q;{\rm screen}\rangle\; ,
\end{equation}
where $|-q;{\rm screen}\rangle$ is the state of the screening charge.  The
vortex pair is actually in the mixed state
\begin{equation}
\rho={1\over\pi}\sum_q|q;\omega_{\rm tot}\rangle\langle q; \omega_{\rm tot}|\;
{}.
\end{equation}
The probability distribution for the scattered vortex will be the incoherent
sum of the probability distributions for the braid eigenstates.

\section{CONCLUSIONS}
This paper has two recurring themes, relating to the non-abelian Aharonov-Bohm
effect and non-abelian statistics.  The first theme is that the non-abelian
Aharonov-Bohm effect provides a natural setting for multi-valued physical
observables.  A particle that travels around a closed path returns to its
starting point as a {\it different} kind of particle with different quantum
numbers.
This means that transition probabilities are not single-valued functions of the
positions and quantum numbers of the particles in the final state.  We have
calculated cross sections that exhibit this multi-valued character.

The second theme is that two particles that are ``indistinguishable'' need not
be the same.  The hallmark of non-abelian statistics is that there can be an
exchange contribution to an amplitude that interferes with the direct
amplitude, even if the two particles that are exchanged are distinct objects
with different quantum numbers.  We have calculated cross sections that include
such exchange effects.

These considerations illuminate some subtle aspects of non-abelian gauge
invariance.  How do they relate to real phenomenology?  There is no firm
evidence that objects that obey non-abelian statistics (called ``nonabelions''
in Ref.~\cite{moore}) exist in nature.  But it is surely conceivable
that
nonabelions will eventually be found, in strongly correlated electron
systems \cite{wilczekwu,moore,wen},
or other frustrated quantum many-body systems.  An important question, then, is
how would such objects be recognized in laboratory experiments?  Much remains
to be done to explore the many-body physics of nonabelions.  Even the problem
of three bodies is not very well understood.

\acknowledgments
We thank Martin Bucher and Kai-Ming Lee for very useful discussions.  This
research was supported in part by DOE Grant No. DE-FG03-92-ER40701.

\figure{Exchange of two vortices.  (a) The paths $\alpha$ and $\beta$ are two
standard paths, both beginning and ending at the same basepoint $x_0$, that are
used to define the flux of two vortices.  (b) When the vortices are
interchanged, these paths are dragged to the new paths
$\beta'=\beta^{-1}\alpha\beta$ and $\alpha'=\beta$.\label{figA}}

\figure{Vortices can be carried along specified paths to the ``Vortex Bureau of
Standard,'' where their flux can be measured.  If the two vortices are carried
along the paths shown in (a), the fluxes are measured to be $a$ and $b$,
respectively.  But if the $b$ vortex goes counterclockwise around the $a$
vortex before voyaging to the Bureau, as in (b), its flux is measured as
$(ab)b(ab)^{-1}$. If the $a$ vortex goes counterclockwise around the $b$ vortex
before voyaging to the Bureau, as in (c), its flux is measured as
$(ab)a(ab)^{-1}$.\label{figB}}

\figure{The charge of a particle can be measured via the Aharonov-Bohm effect
in a double-slit interference experiment.  (a) When a vortex of known flux $b$
is placed between the two slits, the change in the interference pattern
measures $\langle u|D^{(R)}(b)|u\rangle$, where $|u\rangle$ denotes the
internal state of the charged particle, and $(R)$ is the representation
according to which the charged particle transforms.  However, if the charged
particle is itself a vortex with flux $a$, there is a restriction on the
charges that can be measured.  If the $a$ vortex passes through the left slit,
as in (b), it arrives at the screen with flux $a$, and the vortex between the
slits remains in the flux state $b$.  If it passes through the right slit, as
in (c), it arrives at the screen with flux $(ab)a(ab)^{-1}$, and the flux of
the vortex between the slits becomes $(ab)b(ab)^{-1}$. Thus, no interference is
seen if $a$ and $b$ do not commute.  Because interference occurs only when $a$
and $b$ commute, this experiment can measure only the transformation properties
of the charged projectile under the subgroup $N(a)$ that commutes with
$a$.\label{figC}}

\figure{Two paths that contribute to the amplitude for a $b$ vortex propagating
on the background of a fixed $a$ vortex.  If the $b$ vortex passes below the
$a$ vortex, it arrives at its destination with flux $b$; if it passes above the
$a$ vortex, it arrives at its destination with flux $aba^{-1}$.  Thus, these
two paths do not interfere if $a$ and $b$ do not commute.\label{figD}}

\figure{A convention for measuring the flux of a scattered vortex that is
single-valued but discontinuous.  If the vortex is scattered into the upper
half plane ($0<\theta<\pi$), it is carried back to the ``Vortex Bureau of
Standards'' above the scattering center; if the vortex is scattered into the
lower half plane ($-\pi<\theta<0$), it is carried back to the Bureau above the
scattering center.  With this convention, the scattering cross section is
discontinuous at $\theta=0$; the cross section in the ``$k$'' channel at
$\theta=0^+$ matches the cross section in the ``$k+1$'' channel at
$\theta=0^-$.\label{figE}}

\figure{Paths contributing to the amplitude for the propagation of a pair of
vortices.  The initial vortices carry flux taking the values $(12)$ and $(23)$
in $S_3$.  If the vortices braid once as in (b) or twice as in (c), the quantum
numbers of the pair are modified.  But if the vortices braid three times as in
(d), the final quantum numbers match the initial quantum numbers.  Thus, paths
(a) and (d) add coherently in the amplitude, although the two vortices change
places.\label{figF}}


\begin{references}

\bibitem{leinaas}M. Leinaas and J. Myrheim, Nuovo Cimento B {\bf 37} (1977) 1;
G. A. Goldin, R. Menikoff, and D. H. Sharp, J.\ Math.\ Phys.\ {\bf 22} (1981)
1664.
\bibitem{wilczek82}F. Wilczek, Phys.\ Rev.\ Lett.\ {\bf 48} (1982) 1144; {\bf
49} (1982) 957.
\bibitem{fqhe}R. B. Laughlin, Phys.\ Rev.\ Lett.\ {\bf 50} (1983) 1395; F. D.
M. Haldane, Phys.\ Rev.\ Lett.\ {\bf 51} (1983) 605; B. I. Halperin, Phys.\
Rev.\ Lett.\ {\bf 52} (1984) 1583; D. Arovas, J. R. Schrieffer, and F. Wilczek,
Phys.\ Rev.\ Lett.\ {\bf 53} (1984) 722.
\bibitem{goldin85}G. A. Goldin, R. Menikoff, and D. H. Sharp, Phys.\ Rev.\
Lett.\ {\bf 54} (1985) 603.
\bibitem{bala}C. Aneziris, A. P. Balachandran, M. Bourdeau, S. Jo, T. R.
Ramadas, and R. D. Sorkin, Mod.\ Phys.\ Lett.\ A {\bf 4} (1989) 331; T. D.
Imbo, C. S. Imbo, and E. C. G. Sudarshan, Phys.\ Lett.\ B {\bf 234} (1990) 103;
T. D. Imbo and J. March-Russell, Phys.\ Lett.\ B {\bf 252} (1990) 84.
\bibitem{wilczekwu}F. Wilczek and Y.-S. Wu, Phys.\ Rev.\ Lett.\ {\bf 65} (1990)
13.
\bibitem{bucher}M. Bucher, Nucl.\ Phys.\ B {\bf 350} (1991) 163.
\bibitem{mermin}N. D. Mermin, Rev.\ Mod.\ Phys.\ {\bf 51} (1979) 591.
\bibitem{volovik}M. M. Salomaa and G. E. Volovik, Rev.\ Mod.\ Phys.\
{\bf 59} (1987) 533.
\bibitem{schwarz}A. S. Schwarz, Nucl.\ Phys.\ B {\bf 208} (1982) 141.
\bibitem{alfordbenson}M. G. Alford, K. Benson, S. Coleman, J. March-Russell,
and F. Wilczek, Phys.\ Rev.\ Lett.\ {\bf 64} (1990) 1632; Nucl.\ Phys.\ B {\bf
349} (1991) 414.
\bibitem{preskillkrauss}J. Preskill and L. Krauss, Nucl.\ Phys.\ B {\bf 341}
(1990) 50.
\bibitem{bucherlopreskill}M. Bucher, H.-K. Lo, and J. Preskill, Nucl.\ Phys.\ B
{\bf 386} (1992) 3.
\bibitem{bais80}F. A. Bais, Nucl.\ Phys.\ B {\bf 170} (1980) 32.
\bibitem{toulouse}V. Po\'enaru and G. Toulouse, J.\ Phys.\ (Paris) {\bf
38} (1977) 887.
\bibitem{verlinde}E. Verlinde, in {\it Modern Quantum Field Theory}, Bombay
International Colloquium 1990 (World Scientific, Singapore, 1991)
\bibitem{bais92}F. A. Bais. P. van Driel, and M. de Wild Propitius, Phys.\
Lett.\ B {\bf 280} (1992) 63; Nucl.\ Phys.\ B {\bf 393} (1993) 547.
\bibitem{hooft}G. 't Hooft, Comm.\ Math.\ Phys.\ {\bf 117} (1988) 685.
\bibitem{deser}S. Deser and R. Jackiw, Comm.\ Math.\ Phys.\ {\bf 118} (1988)
495.
\bibitem{alfordmrwilczek}M. Alford, J. March-Russell, and F. Wilczek, Nucl.\
Phys.\ B {\bf 337} (1990) 695.
\bibitem{alfordcolemanmr}M. Alford, S. Coleman, and J. March-Russell, Nucl.\
Phys.\ B {\bf 351} (1991) 735.
\bibitem{nonreal}A. P. Balachandran, F. Lizzi, and V. Rogers, Phys.\ Rev.\
Lett.\ {\bf 52} (1984) 1818.
\bibitem{frolich}J. Fr\"ohlich and P.-A. Marchetti, Lett.\ Math.\ Phys.\ {\bf
16} (1988) 347; Comm.\ Math.\ Phys.\ {\bf 121} (1989) 177; Nucl.\ Phys.\ B {\bf
356} (1991) 533.
\bibitem{frolichetc}J. Fr\"ohlich, and F. Gabbiani, Rev.\ Math.\ Phys.\ {\bf 2}
(1991) 251.
\bibitem{orbifold}R. Dijkgraaf, V. Pasquier, and P. Roche, Nucl.\ Phys.\ B
(Proc. Suppl.) 18 {\bf B} (1990) 60.
\bibitem{bantay}P. Bantay, Phys.\ Lett.\ B {\bf 245}
(1990) 477; Lett.\ Math.\ Phys.\ {\bf 22} (1991) 187.
\bibitem{rehren} K.-H. Rehren, Comm.\ Math.\ Phys.\ {\bf 116} (1988) 675.
\bibitem{preskilletc}M. G. Alford, K.-M. Lee, J. March-Russell, and J.
Preskill, Nucl.\ Phys.\ B {\bf 384} (1992) 251.
\bibitem{bucherleepreskill}M. Bucher, K.-M. Lee, and J. Preskill, Nucl.\ Phys.\
B {\bf 386} (1992) 27.
\bibitem{imbo}L. Brekke, H. Dykstra, A. F. Falk, and T. D. Imbo, Phys.\ Lett.\
B {\bf 304} (1993) 127
\bibitem{laidlaw}M. G. G. Laidlaw and C. De Witt-Morette, Phys.\ Rev.\ D {\bf
3} (1971) 1375; L. S. Schulman, J.\ Math.\ Phys.\ {\bf 12} (1971) 304; Y.-S.
Wu, Phys.\ Rev.\ Lett.\ {\bf 52} (1984) 2103.
\bibitem{doplicher}S. Doplicher and J. E. Roberts, Comm.\ Math.\ Phys.\ {\bf
131} (1990) 51.
\bibitem{witdij}R. Dijkgraaf and E. Witten, Comm.\ Math.\ Phys.\ {\bf 129}
(1990) 393.
\bibitem{imbo91}L. Brekke, A. F. Falk, S. J. Hughes, and T. D. Imbo, Phys.\
Lett.\ B {\bf 271} (1991) 73.
\bibitem{balastat}A. P. Balachandran, A. Daughton, Z.-C. Gu, G. Marmo, R. D.
Sorkin, and A. M. Srivastava, Mod.\ Phys.\ Lett.\ A {\bf 5} (1990) 1575.
\bibitem{witten}E. Witten, Comm.\ Math.\ Phys.\ {\bf 121} (1989) 351.
\bibitem{chern}G. Moore and N. Seiberg, ``Lectures on RCFT,'' presented
at the Banff summer school, 1989.
\bibitem{aharonov}Y. Aharonov and D. Bohm, Phys.\ Rev.\ {\bf 115} (1959) 485.
\bibitem{hagen}M. V. Berry, R. G. Chambers, M. D. Large, C. Upstill, and J. C.
Walmsley, Eur.\ J.\ Phys.\ {\bf 1} (1980) 154; C. R. Hagen, Phys.\ Rev.\ D {\bf
41} (1990) 2015.
\bibitem{mrwilczek}J. March-Russell and F. Wilczek, Phys.\ Rev.\ Lett.\
{\bf 61} (1988) 2066.
\bibitem{moore}G. Moore and N. Read, Nucl.\ Phys.\ B {\bf 360} (1991) 362.
\bibitem{wen}X.-G. Wen, Phys.\ Rev.\ Lett.\ {\bf 66} (1991) 802.

\end{references}
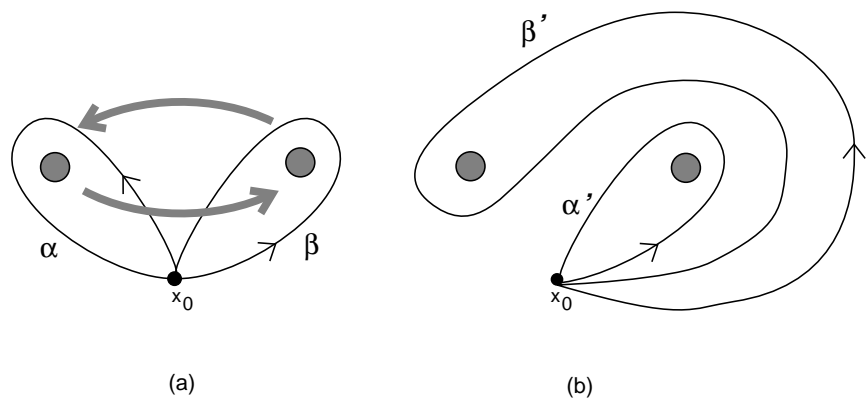
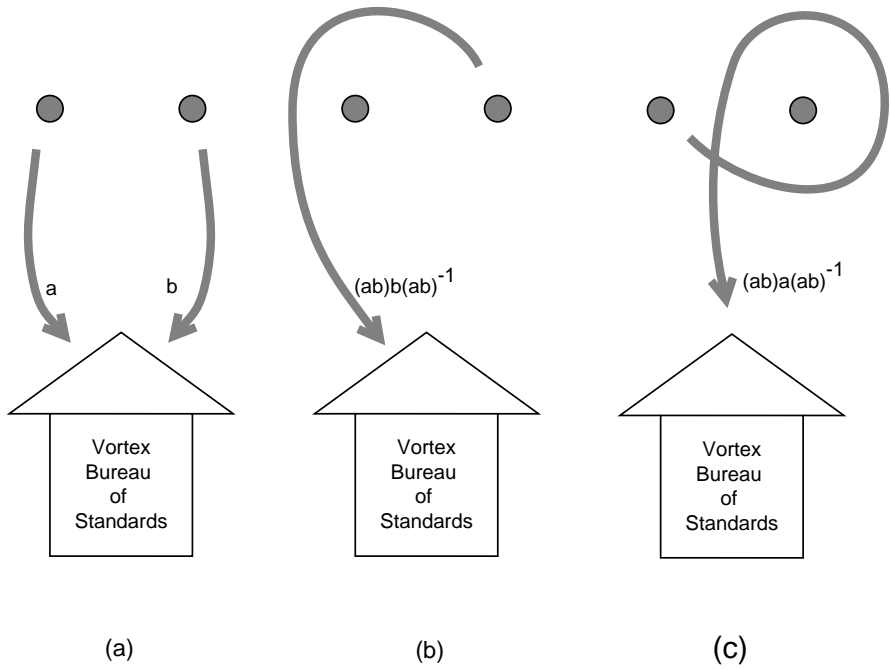
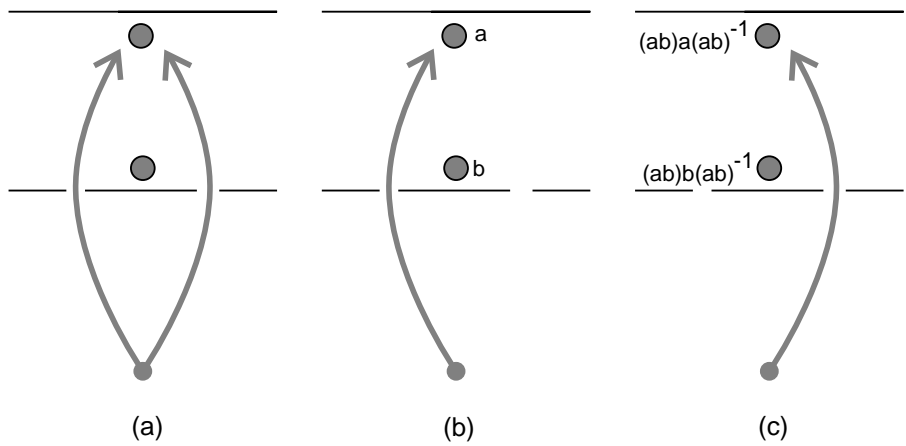
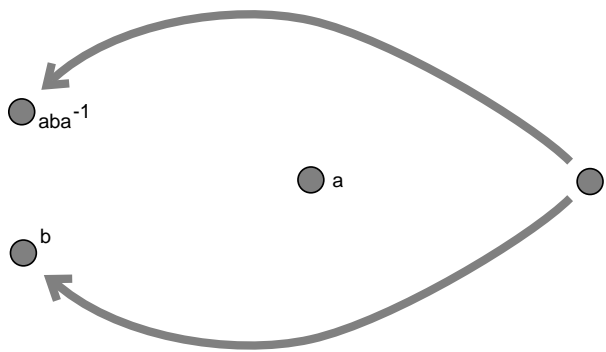
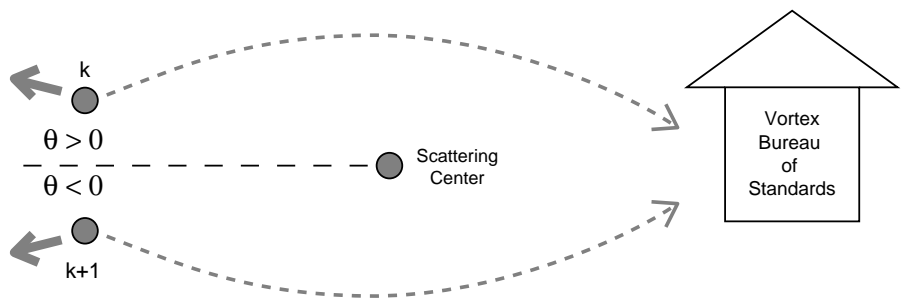
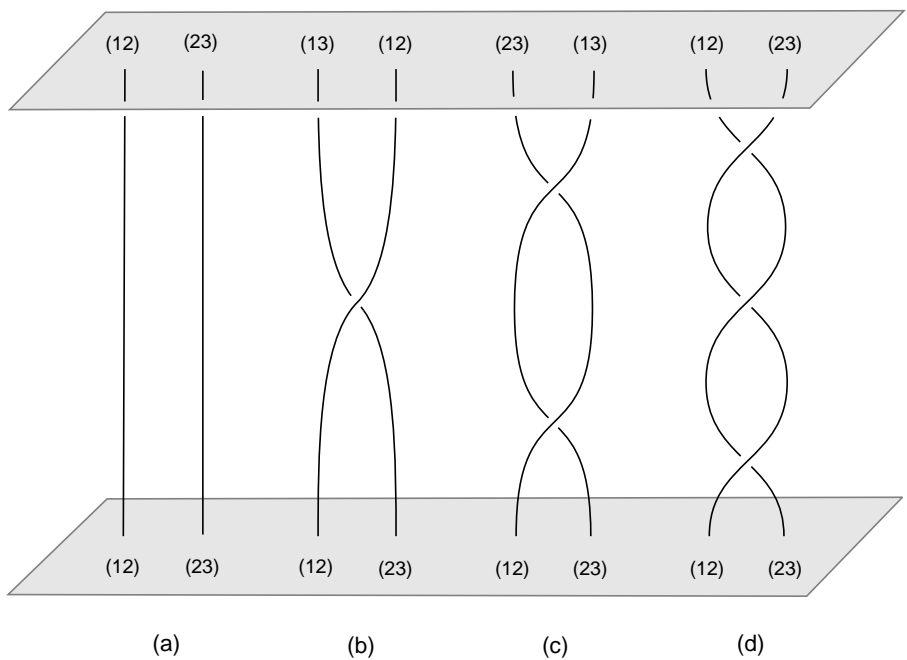
\end{document}